\documentclass[journal=jacsat,manuscript=article,amsmath]{achemso}
\pdfoutput=1
\usepackage{textcomp}
\usepackage[latin9]{inputenc}
\usepackage{color}
\usepackage{booktabs}
\usepackage{amsmath}
\usepackage{amssymb}
\usepackage{graphicx}
\SectionNumbersOn

\makeatletter

\title{Exciton Basis Description of Ultrafast Triplet Separation in Pentacene-(Tetracene)2-Pentacene
Intramolecular Singlet Fission Chromophore.}

\author{Arifa Nazir}

\affiliation{Department of Physics, Indian Institute of Technology Bombay, Powai,
Mumbai 400076, India}

\author{Alok Shukla }

\affiliation{Department of Physics, Indian Institute of Technology Bombay, Powai,
Mumbai 400076, India}

\author{Sumit Mazumdar}

\affiliation{Department of Physics, and the Department of Chemistry, University
of Arizona, Tucson, AZ 85721, USA}

\email{{*}mazumdar@arizona.edu}

\providecommand{\tabularnewline}{\\}
\newcommand{\lyxdot}{.}

\usepackage{color}
\usepackage{ulem}

\usepackage[]{ulem}

\pdfpageheight\paperheight
\pdfpagewidth\paperwidth

\providecommand{\tabularnewline}{\\}
\makeatother

\begin{document}
\begin{abstract}
Precise understanding of the electronic structures of optically dark
triplet-triplet multiexcitons that are the intermediate states in
singlet fission (SF) continues to be a challenge. This is particularly
true for intramolecular singlet fission (iSF) chromophores, that are
oligomers of large monomer molecules. We have performed quantum many-body
calculations of the complete set of excited states relevant to iSF
in Pentacene-(Tetracene)2-Pentacene oligomers, consisting of two terminal
pentacene monomers linked by two teracene monomers. Our computations
use an exciton basis that gives physical pictorial descriptions of
all eigenstates, and are performed over an active space of twenty-eight
monomer molecular orbitals, including configuration interaction with
all relevant quadruple excitations within the active space, thereby
ensuring very high precision. We discuss the many-electron structures
of the optical predominantly intramonomer spin-singlets, intermonomer
charge-transfer excitations, and most importantly, the complete set
of low energy covalent triplet-triplet multiexcitons. We are able
to explain the weak binding energy of the pentacene-tetracene triplet-triplet
eigenstate that is generated following photoexcitation. We explain
the increase in lifetime with increasing numbers of tetracene monomers
of the transient absorption associated with contiguous pentacene-tetracene
triplet-triplet in this family of oligomers. We are consequently able
to give a pictorial description of the triplet separation following
generation of the initial triplet-triplet, leading to a state with
individual triplets occupying only the two pentacene monomers. We
expect many applications of our theoretical approach to triplet separation. 
\end{abstract}
\maketitle

\section{Introduction}

\label{intro}

The optically dark triplet-triplet biexciton $^{1}$(T$_{1}$T$_{1}$),
a bound state of the two lowest triplets T$_{1}$ in $\pi$-conjugated
carbon(C)-based systems, continues to be of strong experimental and
theoretical interest. First discussed in the context of linear $\pi$-conjugated
polyenes, its energetic location below the lowest optically allowed
state \cite{Hudson74a} gave conclusive evidence of strong electron-electron
correlations in linear polyenes \cite{Schulten76a,Ramasesha84c,Tavan87a}.
The latter theoretical results led to more detailed investigations
of electron correlation effects in polyacetylene and related $\pi$-conjugated
polymers \cite{Baeriswyl92a,Soos94a,Ramasesha00a,Barford05a}. More
recently, the $^{1}$(T$_{1}$T$_{1}$) is receiving intense scrutiny
in the context of singlet fission (SF) \cite{Smith13a,Lee13a,Rao17a,Xia17a,Felter19a,Casillas20a,Baldacchino22a},
which refers to the spin-allowed internal conversion of the optical
exciton of a multichromophore $\pi$-conjugated system into $^{1}$(T$_{1}$T$_{1}$),
where the two triplets T$_{1}$ occupy distinct chromophore monomers.
Should the binding energy E$_{b}$ between the individual triplets
T$_{1}$ in the triplet-triplet state $^{1}$(T$_{1}$T$_{1}$) be
small, separation into two free triplets, each of which subsequently
contributes to charge generation, becomes conceivable. While the possibility
of commercial development of organic solar cells with significantly
enhanced quantum efficiency has driven much of the current research
on SF, additional impetus comes from the realization that the spin-quintet
$^{5}$(T$_{1}$T$_{1}$) may be utilized in quantum information devices
\cite{Weiss17a,Tayebjee17a,Smyser20a,Jacobberger22a,Dill23a,Yamauchi24a}.
There is hence significant effort to reach clear understanding of
the dependence of the nature of the spin entanglement and the lifetime
of the $^{1}$(T$_{1}$T$_{1}$) on bonding motifs.

While initial research on SF focused on \textit{inter}molecular SF
(xSF), in which the two triplets of $^{1}$(T$_{1}$T$_{1}$) occupy
nonbonded chromophore monomers, interest has subsequently shifted
to \textit{intra}molecular SF (iSF), that occurs in longer systems
consisting of monomer chromophore molecules linked directly by chemical
bonds or by bridge molecules. Current experimental challenge is to
design iSF chromophores in which there occurs rapid $^{1}$(T$_{1}$T$_{1}$)
generation that overcomes other competing photophysical processes,
followed by rapid dissociation into free triplets, or at least slow
triplet recombination. Meeting both requirements simultaneously is
difficult, as rapid $^{1}$(T$_{1}$T$_{1}$) generation requires
strong intermonomer coupling \cite{Pensack18a,Masoomi-Godarzi20a},
which in turn leads to strong E$_{b}$ that prevents triplet separation
and is responsible for triplet recombination. Indeed, both experiments
and computations find E$_{b}$ in the bulk of iSF compounds to be
substantial and nonconducive to free triplet generation. Understanding
the precise electronic structures of the $^{1}$(T$_{1}$T$_{1}$)
in different families of iSF chromophores, the relationship between
bonding motifs and triplet-triplet binding energy, and in particular,
the paths to triplet separation are all essential for reaching optimized
procedures for controlled triplet pair formation and decay rates.

In what follows we describe our theoretical approach to understand
the photophysics of one family of iSF compounds that has shown great
promise towards overcoming the above challenge. In the process of
understanding this particular family we develop full many-body methods
that can be used to obtain clear quantum mechanical descriptions of
ground and excited state absorptions, as well as triplet-triplet binding
and separation in large iSF molecules that will be useful for the
successful design of future iSF chromophores.

\section{Motivation of work}

\label{PTP} The immediate focus of our theoretical research is to
understand the detailed mechanism of triplet-triplet biexciton generation
and subsequent unusually efficient triplet separation as observed
experimentally in the family of pentacene-(tetracene)$_{n}$-pentacene
oligomers \cite{Pun19a}. These compounds consist of pentacene monomers
linked by multiple tetracene bridge monomers (see Fig.~1(a)), hereafter
\textbf{P-Tn-P} (here \textbf{P} and \textbf{T} represent pentacene
and tetracene monomers, respectively, and \textbf{n} is the number
of tetracenes). We report calculations specifically for \textbf{n
= }2. In what follows we will distinguish between triplet-triplet
configurations in which triplet excitations occupy only pentacene
monomers, $^{1}$(T$_{1[P1]}$T$_{1[P2]})$, and those in which the
triplets occupy both pentacene and tetracene monomers. Triplets in
pentacene-tetracene triplet-triplets can occupy nearest neigbor monomers
or separated more distant monomers. The former will be written as
$^{1}$(T$_{1[P1]}$T$_{1[T1]})$, with the understanding that the
notation refers to the superposition with the configuration related
by spatial symmetry, $^{1}$(T$_{1[P2]}$T$_{1[T2]})$. Similarly
second neighbor pentacene-tetracene triplet-triplets will be written
as $^{1}$(T$_{1[P1]}$T$_{1[T2]})$, with $^{1}$(T$_{1[P2]}$T$_{1[T1]})$
implied. The authors of the experimental work inferred ultrafast generation
of contiguous triplet-triplet pair $^{1}$(T$_{1[P1]}$T$_{1[T1]})$
following photoexcitation of the pentacene singlet optical exciton
\cite{Pun19a} from transient absorption at 1500 nm, an assignment
in agreement with earlier theoretical work \cite{Khan17a}. This excited
state absorption (ESA) was shortlived, and within ultrafast times
was replaced by triplet absorption in the visible from pentacene alone.
According to the authors, the difference in the triplet energies of
\textbf{T} and \textbf{P} monomers ($\sim0.4$ eV) creates an energy
gradient that drives triplet migration and transition to the lowest
energy triplet-pair $^{1}$(T$_{1[P1]}$T$_{1[P2]}$). The latter
has very long lifetime \cite{Pun19a}, as triplet-triplet recombination
is energetically uphill.

\begin{figure}
\centerline{\resizebox{5.5in}{!}{\includegraphics{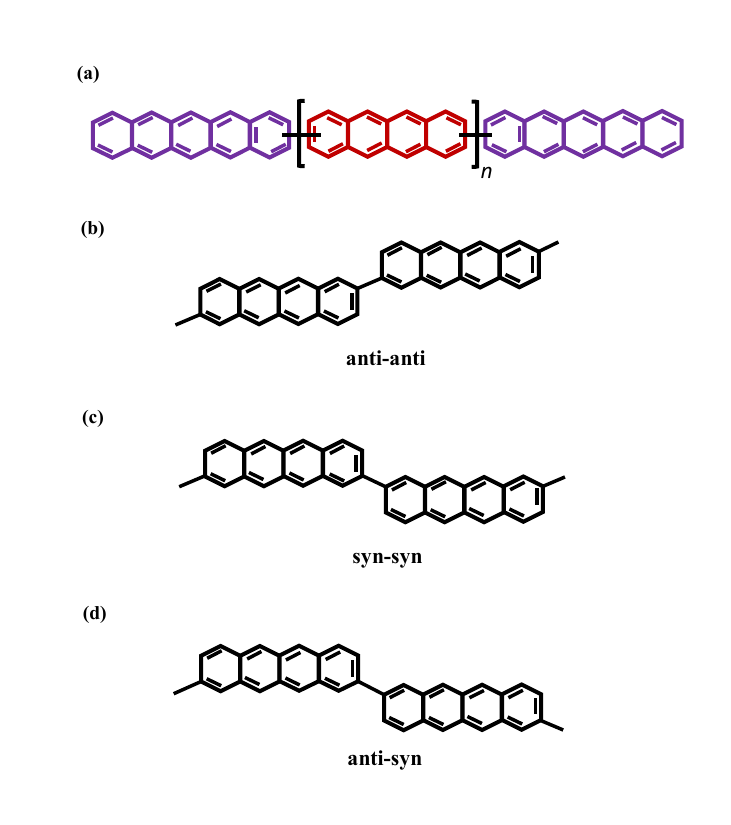}}}
\caption{(Color online) (a) Schematic structure of the \textbf{P-Tn-P} compounds
investigated experimentally \cite{Pun19a}. (b) \textit{anti-anti}
(c) \textit{syn-syn }and\textit{ }(d)\textit{ syn-anti} bonding motifs
investigated theoretically for the \textbf{n=2} molecule.}
\end{figure}

While the experimental results are promising, they raised intriguing
theoretical and experimental questions. In the context of \textbf{n}
= 1, the relevant questions are as follows. First, whether contiguous
triplet-triplet $^{1}$(T$_{1[P1]}$T$_{1[T1]})$ and separated triplet-triplet
$^{1}$(T$_{1[P1]}$T$_{1[P2]}$) indeed constituted completely distinct
eigenstates, in view of the small energy difference of between \textbf{T}
versus \textbf{P} triplets (0.4 eV) relative to the realistic intermonomer
hopping integral ($\sim1.5-2.2$ eV) that would be ``mixing up''
the confgurations. Second, whether triplet migration is indeed behind
the generation of the separated triplet-pair. The latter question
is reasonable, and is based on the earlier demonstration of direct
end-to-end (\textbf{P} to \textbf{P}) CT being the dominant path to
$^{1}$(T$_{1[P1]}$T$_{1[P2]})$ generation in \textbf{P-$\beta$-P}
iSF compounds with $\beta$ = benzene, napthalene and anthracene \cite{Parenti23a}.
In recent theoretical work on \textbf{P-T-P} \cite{Chesler24a} we
have shown explicitly that both these assumptions are correct: contiguous
triplet-triplet generation followed by triplet migration is indeed
the dominant path to separated triplet-pair generation. Direct \textbf{P}
to \textbf{P} CT is the preferred path to iSF only for short bridge
molecules with relatively high bridge monomer absorption threshold,
and is replaced by nearest neighbor CT when the bridge monomer absorption
energy is close to that of the terminal chromophore molecules.

The experimental observations for the ${\bf n}>1$ \textbf{P-Tn-P}
oligomers raise different, though related, questions.

(i) Transient absorption at 1500 nm, associated strictly with the
contiguous triplet-triplet \cite{Khan17a} is also observed in ${\bf n}=2$
and ${\bf n}=3$. The triplet-triplet binding energy E$_{b}$ of the
\textbf{PT} dimer has earlier been calculated to be $\sim0.09$ eV
\cite{Chesler20a}, significantly larger than thermal energy. Straightforward
triplet migration from contiguous triplet-triplet $^{1}$(T$_{1[P1]}$T$_{1[T1]})$
to separated triplet-triplet $^{1}$(T$_{1[P1]}$T$_{1[T2]})$ requires
overcoming E$_{b}$ and appears unlikely. Note that previous attempt
to attain triplet separation in homo-oligomeric polypentacene had
failed \cite{Sanders16d}, an experimental result that is in agreement
with the large E$_{b}$ calculated for bipentacene \cite{Chesler20a}.
What then is the mechanism through which this binding energy is overcome
in ${\bf n}>1$ \textbf{P-Tn-P} ?

(ii) A second interesting experimental observation is that the the
lifetime of the 1500 nm ESA that is used to identify the contiguous
triplet-triplet is unresolvably fast in \textbf{n} =1 , and is 2.6
ps and 5.7 ps in \textbf{n = 2} and \textbf{n = 3}, respectively \cite{Pun19a}.
The increasing lifetimes of the contiguous triplet-triplet in the
longer oligomers is not expected within the simplest triplet exciton
migration scenario. The increase in lifetimes implies that either
the contiguous triplet-triplet lasts longer with increasing n, or
that simplest classification of triplet-triplets as contiguous versus
separated may require modification. This, in turn, raises the question
as to whether distinct pentacene-tetracene and pentacene-pentacene
triplet-triplets continue to persist in ${\bf n}>1$ \textbf{P-Tn-P},
or whether there is some configuration mixing.

A final (albeit minor) issue is whether quantum interference (QI)
effect, which plays a strong role in the photophysics of \textbf{P-$\beta$-P}
with $\beta$ = benzene, napthalene and anthracene \cite{Parenti23a}
but which is virtually nonexistent in \textbf{P-T-P} with ${\bf n}=1$,
plays any role in ${\bf n}>1$ given the tetracene-tetracene linkages.

Resolutions of the above questions necessarily require precise determinations
of electronic structures of the complete set of triplet-triplet multiexciton
wavefunctions in the ${\bf n>1}$ \textbf{P-Tn-P} oligomers. We present
here rigorous many-body description of the spin-singlet excited state
spectrum of \textbf{P-T2-P}, including both the optical singlet and
the multiple optically dark triplet-triplet states. We also present
computational results of ground and excited state absorptions. As
in \textbf{P-T-P} \cite{Chesler24a}, the ESA computation provides
direct evidence for CT-mediated transition from the optical exciton
to the triplet-triplet state generated initially. We show that these
\textbf{n =} 2 calculations, taken together with the experimental
results give pictorial understanding of triplet separation not just
in \textbf{n} = 2, but also in \textbf{n} = 3.

\section{Theoretical approach}

\subsection{Model Hamiltonian and parameterization}

\label{model}

Accurate computations of the triplet-triplet biexciton remains outside
the scope of first principles approaches for systems containing more
than 20 $\pi$-electrons \cite{Kim18a,Rishi23a}. This remains true
in spite of recent progress with quantum chemical method development
\cite{Shao15a,Epifanovsky21a}. The most dominant contributions to
the $^{1}$(T$_{1}$T$_{1}$) eigenstates are two electron-two hole
(2e-2h) excitations from the Hartree-Fock (HF) ground state configuration,
and its correct description absolutely requires incorporation of configuration
interaction (CI) with qudruple 4e-4h excitations. The number of such
4e-4h excitations increases explosively with system size. Computations
over restricted active spaces (RAS) that include only the highest
occupied and lowest unoccupied molecular orbitals (HOMO and LUMO)
of monomers do exist in the literature for small iSF dimers \cite{Korovina16a,Korovina18a},
but are not expected to meet the accuracy requirement for the wavefunctions
in the present case. The Density Matrix Renormalization Group approach
is hard to implement for nonperiodic systems, and to the best of our
knowledge has only been applied to a relatively small dimeric iSF
chromophore, also over the same RAS \cite{Taffet20a}.

Our large-scale computations of triplet-triplet multiexcitons in \textbf{P-T2-P}
are done over an active space of 28 MOs, using the well-tested semiempirical
$\pi$ electron-only Pariser-Parr-Pople (PPP) model Hamiltonian, as
in our previous work on the \textbf{n}=1 oligomer \cite{Chesler24a}.
The Hamiltonian is written

\begin{equation}
H=\sum_{\langle ij\rangle,\sigma}t_{ij}(c_{i\sigma}^{\dagger}c_{j\sigma}+c_{j\sigma}^{\dagger}c_{i\sigma})+U\sum_{i}n_{i\uparrow}n_{i\downarrow}+\frac{1}{2}\sum_{i\neq j}V_{ij}(n_{i}-1)(n_{j}-1)
\end{equation}

Here $c_{i\sigma}^{\dagger}$ creates an electron with spin $\sigma$
on the $p_{z}$ orbital of C-atom $i$, $n_{i\sigma}=c_{i\sigma}^{\dagger}c_{i\sigma}$
is the number of electrons with spin $\sigma$ on atom $i$, and $n_{i}=\sum_{\sigma}n_{i\sigma}$
is the total number of electrons on the atom. The symbol $\langle\rangle$
refers to nearest neigbor C-atoms and $t_{ij}$ are the corresponding
neighbor electron hopping integrals, $U$ the Coulomb repulsion between
two $\pi$ electrons occupying the $p_{z}$ orbital of the same C-atom,
and $V_{ij}$ is long range Coulomb interaction.

Our calculations are for parameters that are the same as in our calculations
for \textbf{P-T-P} \cite{Chesler24a}, which in turn were chosen based
on fittings of optical singlet and triplet states of tetracene and
pentacene monomers \cite{Khan17b,Khan18a}. All
the intramonomer C-C bond lengths (hopping integrals) are taken to
be 1.40 $\mathring{\textrm{A}}$ {} (\textminus 2.4
eV). We have assumed the molecules to be planar for simplicity, with
the interunit bond length 1.46 $\mathring{\textrm{A}}$ and the corresponding
hopping integral $-2.2$ eV, respectively. Monomer rotation effect
can be taken into consideration by reducing the interunit $t_{ij}$
by a multiplicative factor of $cos\theta$, where $\theta$ is the
dihedral angle \cite{Ramasesha90a}. Explicit calculations have confirmed
that physical conclusions are not altered substantively by ignoring
rotation effects \cite{Khan17b} The long range Coulomb interactons
were parametrized as $V_{ij}=U/\kappa\sqrt{1+0.6117R_{ij}^{2}}$,
where $R_{ij}$ is the distance in $\mathring{\textrm{A}}$ between
C-atoms $i$ and $j$ and $\kappa$ is an effective dielectric constant
\cite{Chandross97a}. The onsite Hubbard repulsion $U$ and the dielectric
constant $\kappa$ are taken to be 7.7 eV and $1.3$ based on fitting
monomer energetics \cite{Khan17b,Khan18a}.

\subsection{Diagrammatic exciton basis and computational approach}

Physical understanding of triplet migration requires a pictorial description
of eigenstates that allows clear distinguishing of 1e-1h versus 2e-2h
excitations, as well as intra- versus intermonomer excitations. As
for \textbf{P-T-P} \cite{Chesler24a}, our calculations are done within
the diagrammatic molecular exciton basis \cite{Chandross99a}. We
separate the Hamiltonian into intra- and intermonomer terms (${H=H_{intra}+H_{inter}}$)
where $H_{intra}$ includes all four monomers, and begin with solving
$H_{intra}$ within the Hartree-Fock (HF) approximation to arrive
at single-particle basis states \cite{Sony10a}. We retain 8 frontier
MOs per \textbf{P} monomer (4 bonding and 4 antibonding) and 6 frontier
MOs per \textbf{T} monomer (3 bonding and 3 antibonding), and perform
multiple reference singles and doubles configuration interaction (MRSDCI)
calculatons \cite{Tavan87a} for the complete Hamiltonian of Eq.~1
using the molecular exciton basis to obtain accurate energies and
wavefunctions Supporting Information, hereafter
SI, section I). The MRSDCI procedure incorporates the most dominant
ne-nh excited configurations (n=1-4) that describe each targeted state.
The calculation for each eigenstate is done iteratively, with each
iteration consisting of two stages. In the first stage we perform
a singles and doubles-CI calculation on a basis space of $N_{ref}$
1e-1h and 2e-2h configurations that best describe the targeted eigenstate.
In the second stage we apply the Hamiltonian ($H_{intra}+H_{inter}$)
on the $N_{ref}$ reference configurations. This generates 3e-3h and
4e-4h configurations, of which we retain the most dominant ones to
give the larger Hamiltonian matrix of dimension $N_{total}$. The
larger Hamiltonian matrix usually also contains new 1e-1h and 2e-2h
excited configurations that were not among the original $N_{ref}$
reference configurations, but that are coupled to the 3e-3h and 4e-4h
configurations reached from them. The procedure is repeated with updated
$N_{ref}$ configurations to reach a new larger Hamiltonian with updated
$N_{total}$, until the convergence criterion is reached. Ground and
the excited eigenstate wave functions are therefore treated on the
same footing, while accounting for their multireference character
(see SI Section II). Table S1 in SI
gives the dimensions of the MRSDCI matrices that were needed to reach
convergences within each symmetry subspace. The present calculations
represent the largest correlated-electron calculations of triplet-triplet
multiexcitons in any iSF chromophore to date.

\section{Ground State Absorption}

We calculated the fully correlated ground state wavefunction and wavefunctions
of all excited states coupled to the ground state by the dipole operator,
for all three oligomers in Figs.~1(b)-1(d). Frequency-dependent optical
absorption spectra were obtained from the calculated excited state
energies and dipole couplings, using Lorentzians with uniform linewidths
of 0.05 eV. The absorption spectra, shown in Fig.~2(a), are to be
compared against the experimental spectra in reference \cite{Pun19a}.
The near identical absorption spectra for all three linkages indicates
that quantum interference \cite{Parenti23a} is nearly nonexistent.
\begin{figure}[H]
\centerline{\resizebox{5.5in}{!}{\includegraphics{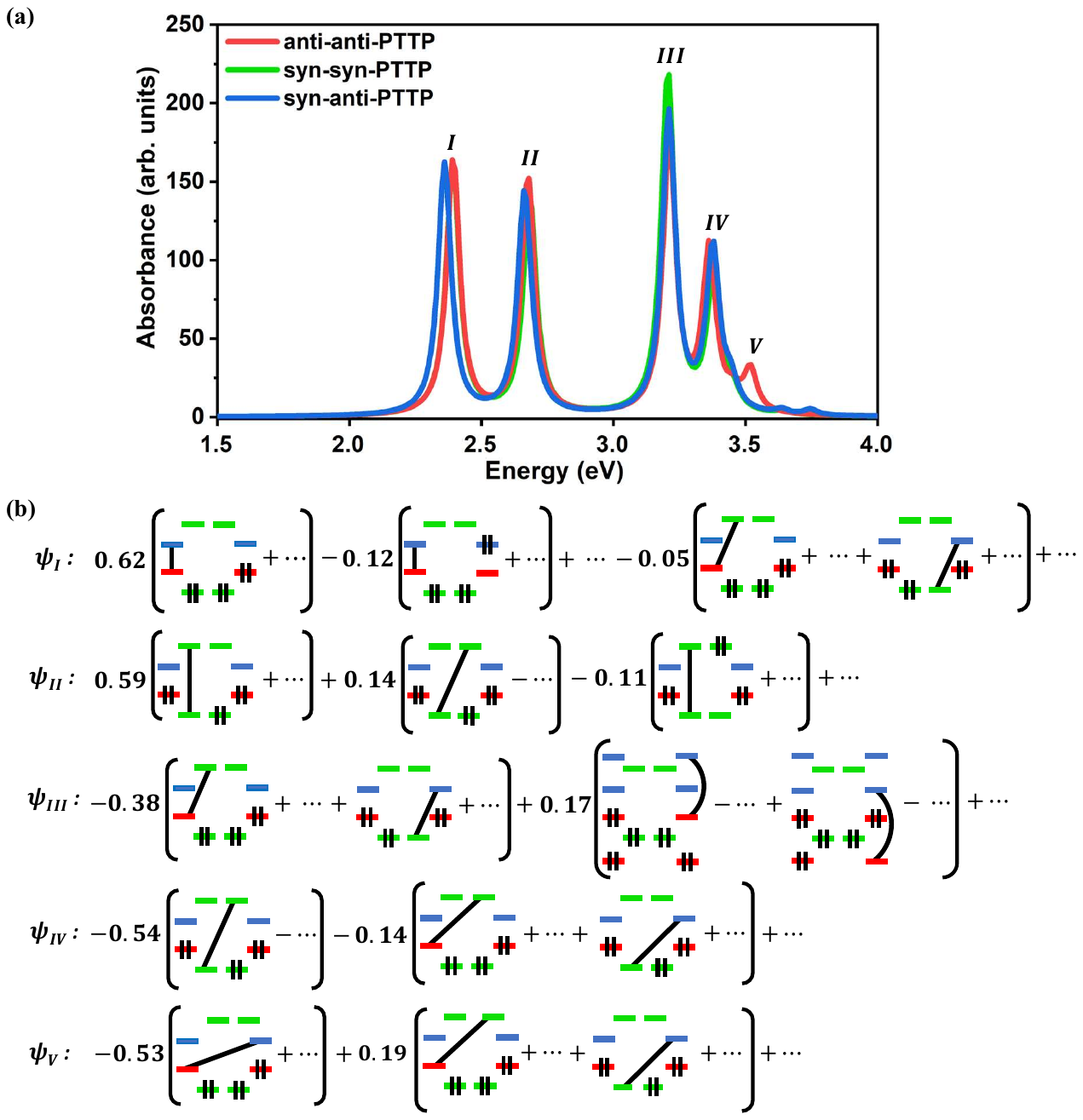}}}
\caption{(a) Calculated ground state optical absorption spectra for the anti-anti,
syn-syn, and syn-anti configurations of \textbf{P-T2-P}. (b) Normalized
exciton basis wavefunctions corresponding to the final states of absorptions
in anti-anti \textbf{P-T2-P}. The outer (inner) pairs of horizontal
bars correspond to the highest occupied bonding and lowest unoccupied
antibonding MOs (HOMOs and LUMOs) of \textbf{P} (\textbf{T}). Not
shown are lower energy bonding (higher energy antibonding) MOs that
are fully occupied (empty). Electron occupancies of MOs are indicated.
Lines connecting MOs are spin-singlet bonds. Ellipses correspond to
additional configurations related by mirror-plane and charge-conjugation
symmetries. See also Figs. S2 and S3 in SI.}
\end{figure}

Fig.~2(b) shows the most dominant exciton basis components (along
with a few select weak components) of the normalized wavefunctions
of the final eigenstates corresponding to each absorption band for
anti-anti PTTP. The corresponding closely related wavefunctions for
syn-syn and syn-anti PTTP are shown in Section III of SI.
The final states of the lowest two absorption bands, labeled I and
II in Fig.~2(a), are dominated by monomer Frenkel excitons, along
with weak \textbf{P1}-to-\textbf{T1} (and \textbf{T2}-to-\textbf{P2})
CT components in I and \textbf{T1}-to-\textbf{T2} CT component in
II. Such intermonomer CT are behind the experimentally observed redshifts
of the absorptions to optical singlet excitons of iSF oligomers relative
to the corresponding monomers \cite{Pun19a}, and are also drivers
of SF (see below). Absorption band III is predominantly due to CT
between \textbf{P1} and \textbf{T1}, with significant mixing with
higher energy pentacene intramonomer excitation. The lowest three
absorption bands also occur in \textbf{P}-\textbf{T}-\textbf{P}, and
the final states there are very similar \cite{Chesler24a}. Absorption
band IV is predominantly \textbf{T1}-to-\textbf{T2} CT, with weak
but nonzero admixing with second neighbor \textbf{P1}-to-\textbf{T2}
CT. The weak absorpion band V is predominantly due to direct long-range
CT between \textbf{P1} and \textbf{P2}. Such long-range CT requires
strong constructive quantum interference that can occur only with
anti-anti linkage \cite{Parenti23a}.

\section{CT-mediated $^{1}$(T$_{1}$T$_{1}$) generation}

Direct evidence for CT-mediated $^{1}$(T$_{1}$T$_{1}$) generation
in the iSF compounds \textbf{P-$\beta$-P}, $\beta$ = benzene, napththalene
and anthracene was seen from the very strong quantum interference
effects on the $^{1}$(T$_{1}$T$_{1}$) generation time in these.
Triple-triplet generation time is significantly faster for anti-linkage
than for syn-linkage, and there is one-to-one correspondence between
this generation time and the strength of CT absorption, observed both
experimentally and computationally \cite{Parenti23a}. As seen in
Fig.~2(a), however, interference effect on the absorption spectra
are weak here, in that the strengths of absorption bands III and IV
are independent of bonding motifs. This is not unexpected, as quantum
interference is a consequence of antiferromagnetic spin correlation
driven by the Hubbard $U$ in Eq.~(1), but such correlation in effectively
one-dimensional systems is short range. The very strong CT absorptions
with all three linkages in Fig.~2(a) suggest equally strong role
of CT for all linkage motifs. The exciton basis computational approach
nevertheless allows demonstration of CT-mediation of $^{1}$(T$_{1}$T$_{1}$)
generation in \textbf{P-T2-P}, as we discuss below.

The authors of a recent experimental work on xSF in pentacene crystals
successfully demonstrated CT-mediated $^{1}$(T$_{1}$T$_{1}$) generation
from careful transient absorption measurements \cite{Neef23a}. According
to the authors CT-mediated SF requires configuration mixing of the
singlet Frenkel exciton and CT configuration even in the absence of
substantial nuclear rearrangement, which in turn leads to mixing of
the CT configuration and the $^{1}$(T$_{1}$T$_{1}$) configuration
in the dipole-coupled virtual excited state from which the triplet-triplet
is generated. In Fig.~3 we have given a schematic respresentation
of the proposed mechanism. The schematic suggests that the eigenstate
reached by ESA from $\Psi_{I}$ in Fig.~2(b) should have weak contribution
from a triplet-triplet configuration. In Fig.~4 we have shown the
calculated final eigenstates that are reached by ESA from the lowest
optical singlet for all three bonding motifs of Fig.~1. In all cases
the last many-electron configuration shown is $^{1}$(T$_{1[P1]}$T$_{1[T1]}$).
\begin{figure}[H]
\centerline{\resizebox{4.5in}{!}{\includegraphics{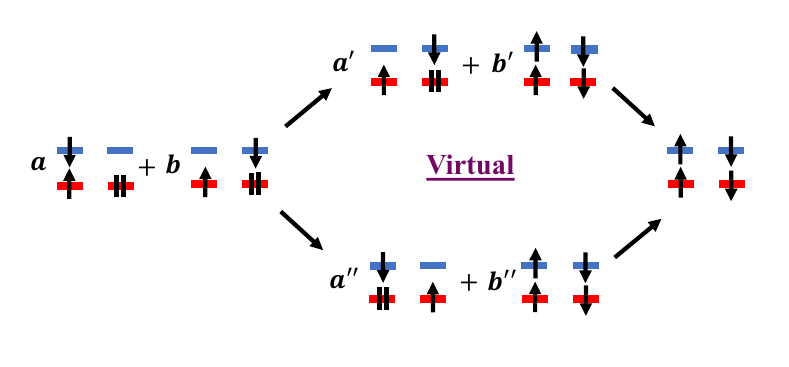}}}
\caption{Schematic of CT-mediated SF mechanism in a chromophore dimer. Only
the HOMO and LUMO of each monomer and their electronic occupancies
are shown. At the very left is the optical singlet, which is a superposition
of Frenkel exciton and CT configuration. It is understood that $b<<a$.
Real or virtual optical excitation from this will necessarily lead
to the intermediate mixed CT and triplet-triplet state, again with
$b^{\prime}<<a^{\prime}$. The final state is the triplet-triplet
multiexciton. }
\end{figure}

\begin{figure}[H]
\centerline{\resizebox{5.5in}{!}{\includegraphics{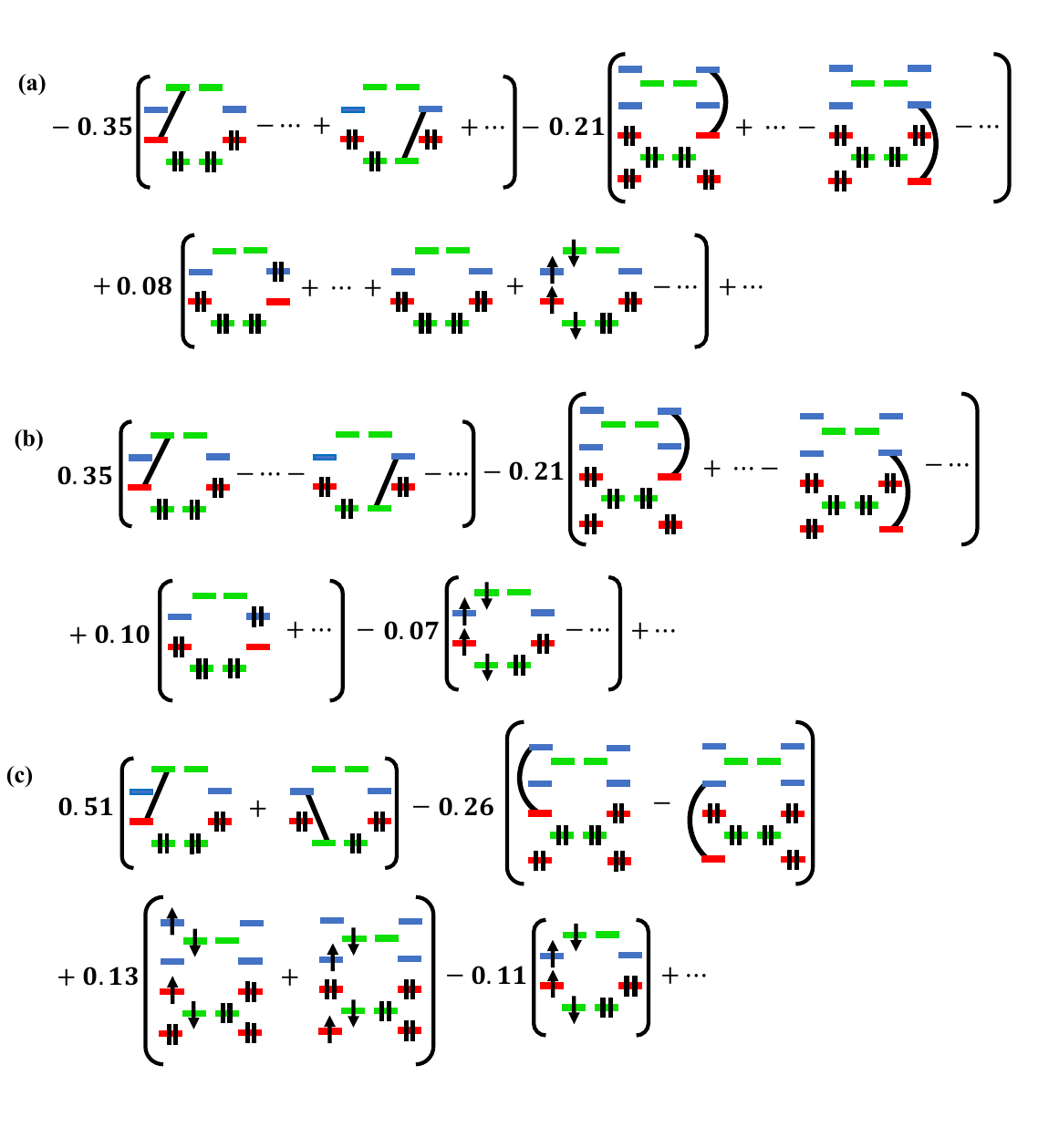}}}
\caption{Normalized wavefunctions of eigenstates to which ESAs from the optical
spin singlet state are the strongest, for (a) anti-anti, (b) syn-syn,
and (c) syn-anti connectivity. In agreement with the schematic of
Fig.~3(b) in each case the eigenstate has a weak but nonvanishing
$^{1}$(T$_{1[P1]}$T$_{1[T1]}$) component.}
\end{figure}

\section{Triplet-Triplet States.}

\subsection{Wavefunctions and Energies.}

In Fig.~5 we show the dominant exciton basis configurations that
constitute the four distinct classes of triplet-triplet eigenstates
in anti-anti \textbf{P-T2-P}. We emphasize that our near-exact wavefunctions
show no mixing between pentacene-pentacene and pentacene-tetracene
triplet-triplet configurations, or between the latter and tetracene-tetracene
triplet-triplet. Table 1 gives the corresponding calculated energies,
to be compared against the calculated energy of 2.39 eV of the lowest
singlet exciton ($\Psi_{I}$ in Fig.~2(a)). The corresponding data
for syn-syn and syn-anti bonding motifs are given in sections VI-VIII
of SI. The lowest energy eigenstate with triplets
localized on the terminal pentacene monomers is nearly identical to
that in \textbf{P-T-P}, while the highest energy eigenstate with triplets
localized on tetracenes is absent there. More interesting are the
pairs of doubly degenerate triplet-triplets occupying pentacene and
tetracene monomers, shown in Figs.~5(b) and (c). These wavefunctions
are not simply contiguous or separated triplet-triplets, but their
superpositions. We have labeled the pair of states with larger contribution
from neighboring monomers as $^{1}$(T$_{1[P1]}$T$_{1[T1][T2]}$)
and the pair of states with larger contribution from second neighbor
monomers as $^{1}$(T$_{1[P1]}$T$_{1[T2][T1]}$), respectively. The
similar natures of the superpositions of the triplet-triplet configurations
in Figs.~5(b) and (c) explains the very small energy difference between
the corresponsing eigenstates (Table I), which is nearly an order
of magnitude smaller than the previously calculated binding energy
of $\sim$ 0.09 eV of the pentacene-tetracene dimer triplet-triplet
\cite{Chesler20a} as well as thermal energy.

\begin{figure}
\centerline{\resizebox{5.5in}{!}{\includegraphics{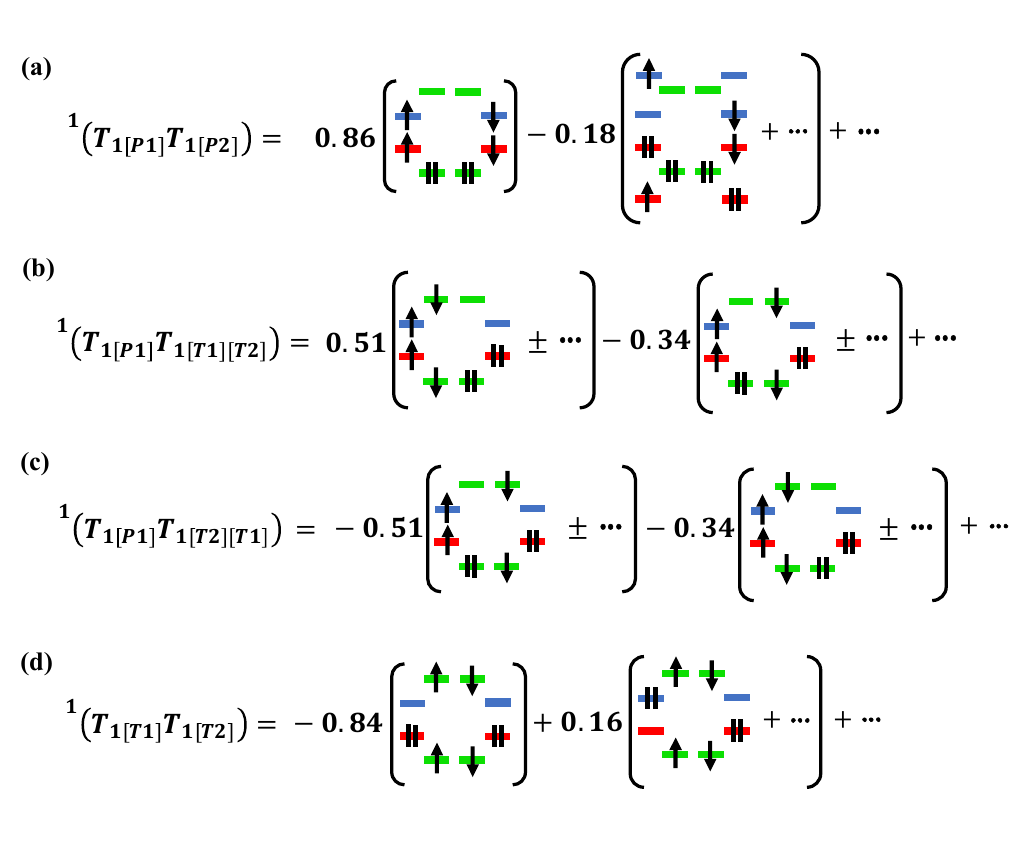}}}
\caption{Dominant components of normalized triplet-triplet eigenstates of anti-anti
\textbf{P-T2-P} (a) $^{1}$(T$_{1[P1]}$T$_{1[P2]}$); (b) $^{1}$(T$_{1[P1]}$T$_{1[T1][T2]}$),
superposition of $^{1}$(T$_{1[P1]}$T$_{1[T1]}$) and $^{1}$(T$_{1[P1]}$T$_{1[T2]}$);
(c) $^{1}$(T$_{1[P1]}$T$_{1[T2][T1]}$), also a superposition of
the same basic triplet-triplet configurations as in (b) but with the
relative weights reversed; and (d) $^{1}$(T$_{1[T1]}$T$_{1[T2]}$)
states in anti-anti \textbf{P-T2-P}. Eigenstates (b) and (c) are each
doubly degenerate.}
\end{figure}

\begin{table}
\caption{Calculated energies (in eV) for $^{1}$(T$_{1[P1]}$T$_{1[P2]}$),
$^{1}$(T$_{1[P1]}$T$_{1[T1][T2]}$), $^{1}$(T$_{1[P1]}$T$_{1[T2][T1]}$)
and $^{1}$(T$_{1[T1]}$T$_{1[T2]}$) for anti-anti\textbf{ P-T2-P}.
The calculated energy of pentacene singlet exciton is 2.39 eV.}
\begin{tabular}{cc}
\toprule 
State  & Energy (eV)\tabularnewline
\midrule
\midrule 
$^{1}$(T$_{1[P1]}$T$_{1[P2]}$)  & 2.09\tabularnewline
\midrule 
$^{1}$(T$_{1[P1]}$T$_{1[T1][T2]}$)  & 2.48\tabularnewline
\midrule 
$^{1}$(T$_{1[P1]}$T$_{1[T2][T1]}$)  & 2.49\tabularnewline
\midrule 
$^{1}$(T$_{1[T1]}$T$_{1[T2]}$)  & 2.77\tabularnewline
\bottomrule
\end{tabular}
\end{table}

\subsection{Excited State Absorptions from pentacene-tetracene triplet-triplets}

The significant contributions by the contiguous pentacene-tetracene
triplet-triplet $^{1}$(T$_{1[P1]}$T$_{1[T1]}$) to both pairs of
triplet-triplet eigenstates (b) and (c) suggest that both will exhibit
ESA to the same singlet CT state \cite{Khan17a}, albeit with weaker
intensity from states (c) with lower relative weight of the contiguous
triplet-triplet. We have calculated ESAs from both classes of eigenstates.
The calculated spectra are shown in Fig.~6. In agreement with previous
calculations on pentacene dimers \cite{Khan17a}, the final state
of the calculated ESA at $\sim1698$ nm from both triplet-triplets
is the same CT state $\Psi_{III}$ in Fig.~2(b). We also find a second
weaker ESA to an intramonomer excited eigenstates at a slightly shorter
wavelength. A broad band at $\sim1250$ nm is indeed visible in the
experimental transition absorption spectrum (see Fig.~8 in Supplementary
Material of reference \cite{Pun19a}), although it was not identified
as a distinct transient absorption by the experimentalists.

\begin{figure}[H]
\centerline{\resizebox{4.5in}{!}{\includegraphics{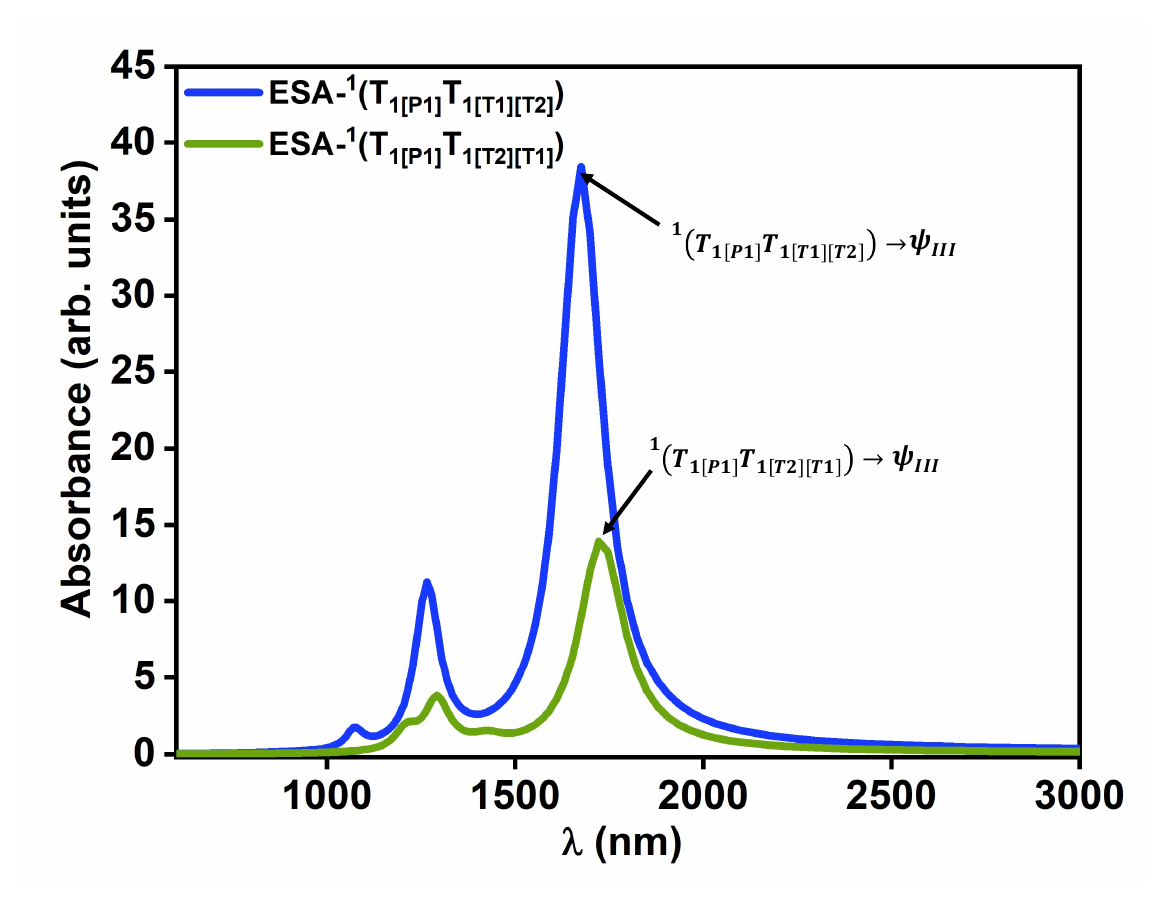}}}
\caption{(a) Calculated excited state absorption spectra from the triplet-triplets
$^{1}$(T$_{1[P1]}$T$_{1[T1][T2]}$) and $^{1}$(T$_{1[P1]}$T$_{1[T2][T1]}$)
for the anti-anti configuration of \textbf{P-T2-P}.}
\end{figure}

\section{Discussion and Conclusion}

In summary, we have performed high level configuration interaction
calculations of the electronic structures of optically allowed singlet
and optically dark triplet-triplet eigenstates of \textbf{P-T2-P}.
We have also performed calculations of ground and excited state absorption
spectra, and have obtained in both cases pictorial descriptions of
the final eigenstates to which these absorptions occur, in order to
understand why the ``cleft'' mechanism \cite{Pun19a} allows generation
of weakly bound triplet-triplets with very long lifetimes. In Fig.~7
we present a schematic of the mechanism of initial pentacene-tetracene
triplet-triplet generation and separation, followed by the final generation
of the pentacene-pentacene triplet-triplet. %\includegraphics[scale=0.15]{fig.6.pdf}
\begin{figure}[H]
\centerline{\resizebox{6.5in}{!}{\includegraphics{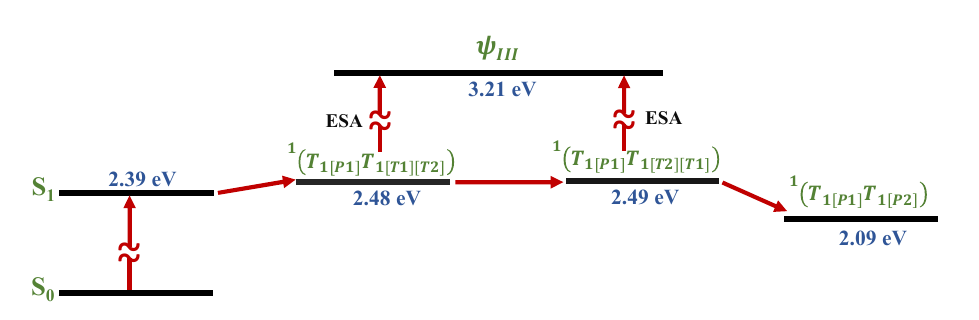}}}
\caption{Schematic of the mechanism of triplet-triplet generation and separation
in \textbf{P-T2-P} (see text).}
\end{figure}

As in the case of \textbf{P-T-P}, we found that pentacene-tetracene
and pentacene-pentacene triplet-triplets constitute distinct eigenstates,
in spite of the intermolecular electron hopping integral being significantly
larger than the difference between tetracene and pentacene triplet
energies. The significantly lower energy of $^{1}$(T$_{1[P1]}$T$_{1[P2]}$)
confers long lifetime to this state, as recombination would be forbiddingly
uphill. More importantly though, this energy gradient leading to the
final triplet-triplet eigenstate influences the composition of the
intermediate pentacene-tetracene triplet-triplet eigenstates in \textbf{P-T2-P};
these can no longer be classified as contiguous versus separated.
Doubly degenerate pairs of pentacene-tetracene triplet-triplet eigenstates
are each quantum mechanical superpositions of configurations with
triplets occupying nearest neighbor monomers as well as second neighbor
monomers. The eigenstates with larger relative weight of the configurations
with increased triplet-triplet distance occur at a very slightly higher
energy that is thermally accessible. Triplet-triplet binding energy
is thus insignificant.

We also reached precise understanding of the increase with increasing
\textbf{n} of the apparent lifetimes of the ``contiguous'' pentacene-tetracene
triplet-triplet, as obtained from the duration of the transient absorption
in the IR \cite{Pun19a}. As shown from computations in Fig.~6 and
is indicated in the schematic of Fig.~7, this transient absorption
due to CT between neighboring triplets \cite{Khan17a} will continue
to occur even as triplet separation proceeds. There are overall four
pentacene-tetracene triplet-triplet eigenstates after taking into
consideration spatial symmetries in \textbf{n = }2. It is to be anticipated
that there will be nine such pentacene-tetracene triplet-triplet eigenstates
with different relative weights of configurations with contiguous
triplets in \textbf{n= }3. The larger number of such eigenstates ensures
greater width of their energy eigenspectrum, and hence longer duration
of the near IR transient absorption to the \textbf{n = }3 equivalent
of $\Psi_{I}$ in Fig.~2(b), as is observed experimentally.

In conclusion, high order CI calculations within the PPP Hamiltonian
using the exciton basis not only allow understanding the electronic
structures of triplet-triplet excitations, but with creative implementation
they will allow understanding in detail photophysical processes that
lead to their unbinding. Fast generation of the initial triplet-triplet
will continue to be one central goal of SF research, and it is likely
that this will lead to even greater focus on iSF (as opposed to xSF)
chromophores, precisely systems in which triplet-triplet binding energy
is relatively large. Clear understanding of the dependence of charge-transfer
and/or triplet exciton migration on bonding motifs then becomes imperative.
It is for instance conceivable that bridge monomers with the lowest
absorption polarized along the lengths of the molecules, as opposed
to transverse to their lengths as is true for acene bridges, will
lead to more efficient charge-transfer. This can often be achieved
with substitution of carbon atoms with heteroatoms containing lone-pairs
of electrons. Greater charge-transfer can conceivably be achieved
also by increasing the effective dimensionality of the experimental
system. Effective PPP-CI calculations can guide focused experimental
search for the ``ideal'' iSF chromophores. We anticipate many future
applications of our theoretical approach.
\begin{suppinfo}
A detailed decription of MRSDCI methodology employed
in this work, along with the normalized exciton basis wavefunctions
corresponding to the final states of absorptions and triplet-triplet
states in syn-syn and syn-anti PT2P, is provided in the Supporting
Information.
\end{suppinfo}
\begin{acknowledgement}
A. S. and S. M. acknowledge fruitful discussions with L. M. Campos
(Columbia University) and M. Sfeir (CUNY). Work at University of Arizona
Tucson was partially supported by NSF Grant No. NSF-DMR-2301372. Some
of the calculations were performed using high performance computing
resources maintained by the University of Arizona Research Technologies
department and supported by the University of Arizona Technology and
Research Initiative Fund, University Information Technology Services,
and Research, Innovation, and Impact.
\end{acknowledgement}
 \bibliography{Fission-NewRefs}

\end{document}

\title{Supporting Information\\
}
\title{Exciton Basis Description of Ultrafast Triplet Separation in Pentacene-(Tetracene)2-Pentacene
Intramolecular Singlet Fission Chromophore.}
\author{{\normalsize Arifa Nazir$^{1}$ }}
\author{Alok Shukla$^{1}$}
\author{Sumit Mazumdar$^{2}$}
\affiliation{$^{1}$Department of Physics, Indian Institute of Technology Bombay,
Powai, Mumbai 400076, India}
\affiliation{$^{2}$Department of Physics and the Department of Chemistry, University
of Arizona, Tucson, AZ 85721, USA}
\maketitle

\section{Molecular exciton basis and active space}

Our objective is to develop intuitive, visual representations of eigenstates.
This requires differentiating excitations that remain localized within
individual monomers from those that spread over multiple units. To
achieve this, we express the PPP Hamiltonian as $H$= $H_{intra}$+
$H_{inter}$ , where it is decomposed into two components:
\begin{enumerate}
\item Intra-unit interaction: $H_{intra}=\sum_{\mu}H_{intra}^{\mu}$ represents
the sum of PPP Hamiltonians for each molecular unit $\mu$ (which
includes the two pentacene monomers and two tetracene bridge molecules)
\item Inter-unit interactions: $H_{inter}=\dfrac{1}{2}\sum_{\nu\neq\mu}H_{inter}^{\nu\mu}$
accounts for hopping ($t_{ij}$) and Coulomb interactions ($V_{ij}$)
between carbon atoms \ensuremath{i} and \ensuremath{j} in different
molecular units. 
\end{enumerate}
The calculations proceed in multiple steps. First, the intra-unit
Hamiltonians ($H_{intra}^{\mu}$) are solved at the Hartree-Fock (HF)
level to generate molecular orbitals (MOs) localized on each unit.
The configuration interaction (CI) matrix is then constructed by iteratively
applying $H$= $H_{intra}$+ $H_{inter}$ to the most probable configurations
that contribute to a given excited state. These configurations, spanning
1e-1h to 4e-4h excitations from the HF ground state, include both
neutral states, where the number of \textgreek{\textpi}-electrons
matches the carbon atom count in each unit, and ionic states, where
the units acquire opposite positive and negative charges.

In our MRSDCI calculations, we concentrate on a strategically chosen
subset of molecular orbitals (MOs) centered around the chemical potential
rather than employing the complete orbital set. To simplify the configuration
interaction process, we \textquotedbl freeze\textquotedbl{} the lowest-energy
bonding MOs, meaning that we do not allow excitations from these orbitals
- and we also remove the highest-energy orbitals, which are related
to them by charge-conjugation symmetry, from the active space. Consequently,
our calculations are performed using active spaces of 28 MOs, divided
into 14 bonding and an equal number of antibonding orbitals as shown
in Fig. S1. Notably, these represent the largest active spaces ever
utilized for \ensuremath{\mid}TT\textrangle{} state calculations.

\begin{figure}[H]
\centering{}\includegraphics[scale=0.2]{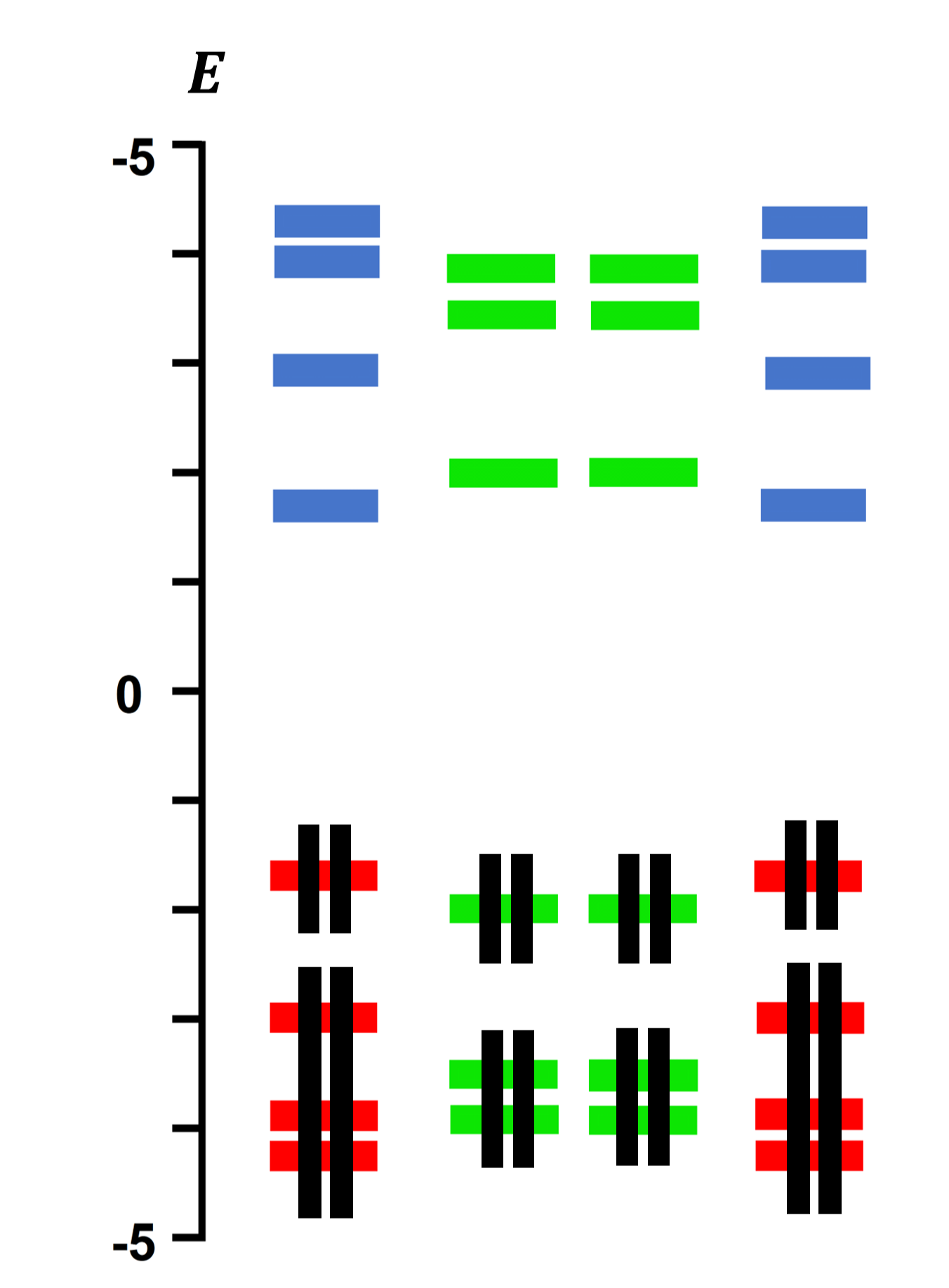}\caption{The energy level diagram for P-T-T-P, as determined at the Hartree-Fock
(HF) level, shows a total of 28 molecular orbitals (MOs) in active
space. The bonding MOs of pentacene are depicted in red, the antibonding
MOs in blue, and the MOs of tetracene in green. Electron occupancy
is indicated by vertical black lines. Every bonding MO is completely
occupied in the HF ground state, whereas every antibonding MO is unoccupied.}
\end{figure}

\section{Multireference Singles and Doubles CI}

To accurately compute $^{m}$(T$_{1}$T$_{1}$) energies and wavefunctions,
we implement an MRSDCI framework that individually targets each eigenstate.
In our approach, configuration interaction (CI) is performed with
excitations up to quadruple order starting from the Hartree-Fock (HF)
ground state. Unlike conventional QCI, which exhaustively includes
all quadruple excitations, we selectively incorporate only the most
relevant \ensuremath{n}-electron \ensuremath{n}-hole (n=1 to 4) excitations
that significantly contribute to each eigenstate. For each eigenstate,
the process starts with a comprehensive singles and doubles-CI calculation
that spans all the single and double excitations, now including the
effects of \ensuremath{H}$_{inter}$(i.e., \ensuremath{H}$_{inter}$\ensuremath{\neq}
0). During this step, we discard any single and double excitations
whose contributions fall below a certain threshold. The remaining
configurations form our initial reference set, denoted as $N_{ref}$.
Next, we perform further CI calculations by including additional double
excitations derived from the $N_{ref}$ configurations. This step
effectively brings in the most important triple and quadruple excitations.
These higher-order excitations can interact strongly with new single
and double excitations that were not part of the original reference
set. By iteratively updating $N_{ref}$ to include these newly significant
configurations, we ensure that all dominant excitations: single, double,
triple, and quadruple with coefficients of 0.1 or greater are incorporated
for each targeted eigenstate. The resulting Hamiltonian, incorporating
all these contributions has a total size of $N_{total}$. Table S1
summarizes the values of $N_{total}$ and $N_{ref}$ for various eigenstates
computed with this method.

\begin{table}[H]
\caption{N$_{ref}$ and N$_{total}$ obtained from MRSDCI calculation in\textit{
anti-anti}, \textit{syn-syn} and \textit{syn-anti} \textbf{PT2P}}

\centering{}%
\begin{tabular}{|c|cc|cc|c|c|}
\hline 
\multicolumn{3}{|c|}{anti-anti} & \multicolumn{2}{c|}{syn-syn} & \multicolumn{2}{c|}{syn-anti}\tabularnewline
\hline 
\hline 
states & N$_{ref}$ & N$_{total}$ & N$_{ref}$ & N$_{total}$ & N$_{ref}$ & N$_{total}$\tabularnewline
\hline 
\ensuremath{S}$_{0}$\ensuremath{S}$_{1}$ & \multirow{2}{*}{53} & \multirow{2}{*}{2588877} & \multirow{2}{*}{59} & \multirow{2}{*}{3065121} & \multirow{2}{*}{70} & \multirow{2}{*}{3564528}\tabularnewline
\cline{1-1}
gs CT &  &  &  &  &  & \tabularnewline
\hline 
ESA CT & 65 & 3299157 & 87 & 4771823 & 82 & 4388608\tabularnewline
\hline 
$^{1}$(T$_{1}$T$_{1}$) & 53 & 2588877 & 59 & 3065121 & 70 & 3564528\tabularnewline
\hline 
\end{tabular}
\end{table}

\section{Eigenstates that contribute to ground-state absorption.}

\begin{figure}[H]
\centering
\begin{centering}
\includegraphics[scale=0.22]{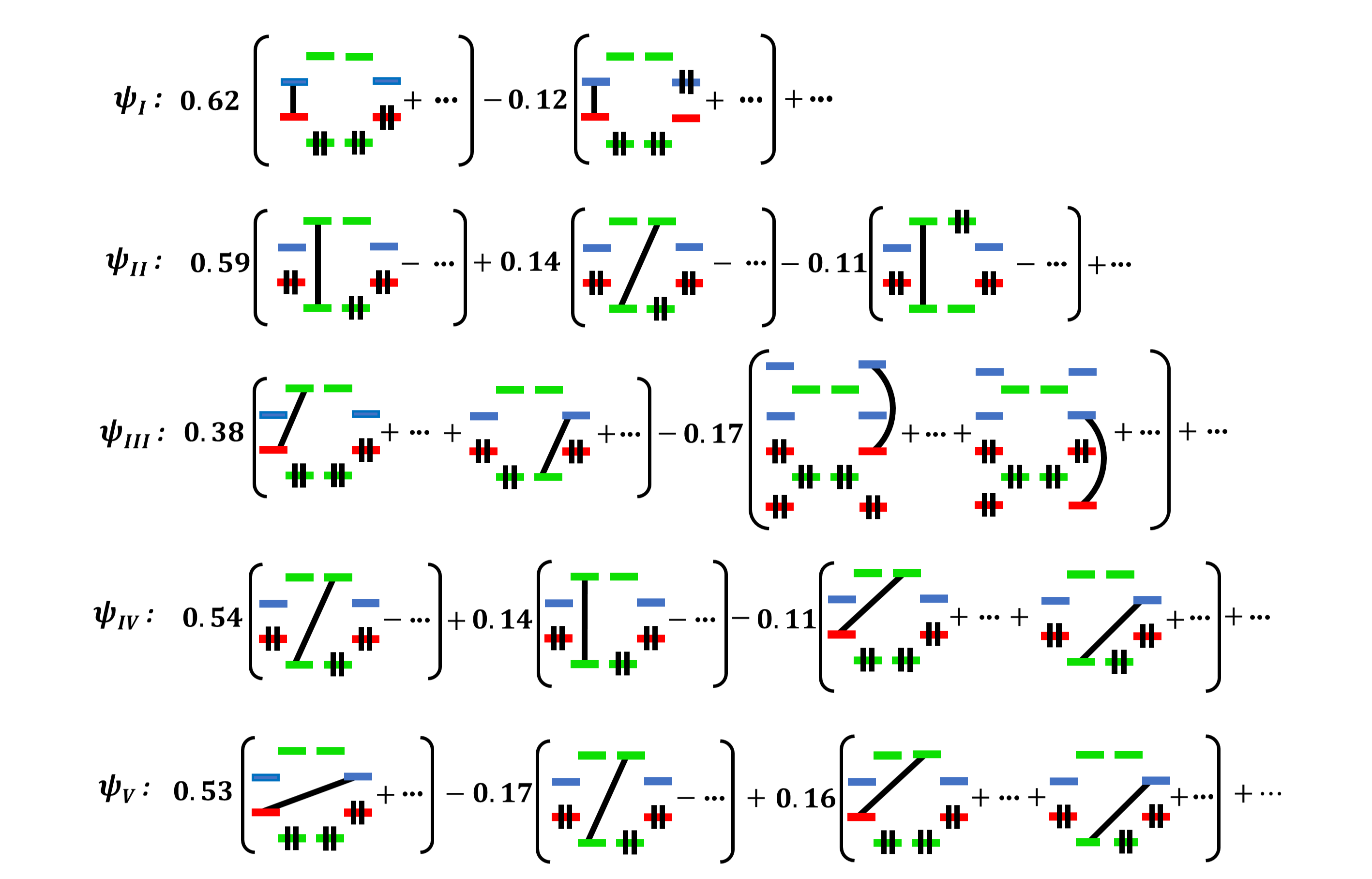}
\par\end{centering}
\caption{Normalized exciton basis wavefunctions corresponding to the final
states of absorptions in syn-syn \textbf{P-T-T-P}. The outer (inner)
pairs of horizontal bars correspond to the highest occupied bonding
and lowest unoccupied antibondingMOs (HOMOs and LUMOs) of P (T). Not
shown are lower energy bonding (higher energy antibonding) MOs that
are fully occupied (empty). Electron occupancies of MOs are indicated.
Lines and curves connecting MOs are spin-singlet bonds.Ellipses correspond
to additional configurations related bymirror-plane and charge-conjugation
symmetries.}

\end{figure}

\begin{figure}[H]
\centering{}\includegraphics[scale=0.22]{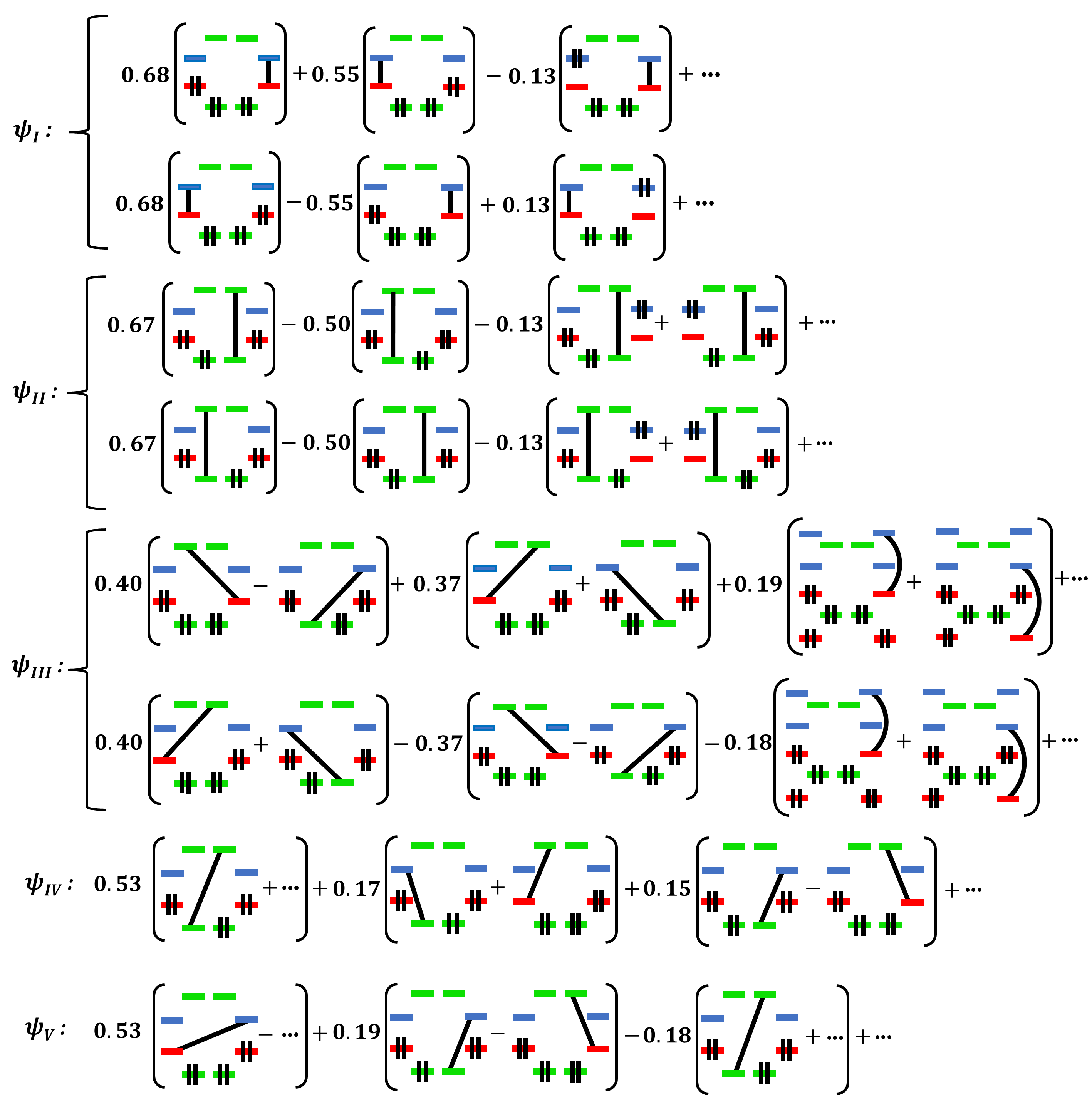}\caption{Normalized exciton basis wavefunctions corresponding to the final
states of absorptions in syn-anti \textbf{P-T-T-P}. Here first three
peaks (I, II, III) are doubly degenerate, hence two eigenstates correspond
to these peaks. The outer (inner) pairs of horizontal bars correspond
to the highest occupied bonding and lowest unoccupied antibonding
MOs (HOMOs and LUMOs) of P (T). Not shown are lower energy bonding
(higher energy antibonding) MOs that are fully occupied (empty). Electron
occupancies of MOs are indicated. Lines and curves connecting MOs
are spin-singlet bonds. Ellipses correspond to additional configurations
related by mirror-plane and charge-conjugation symmetries.}
\end{figure}

\section{Charge transfer states in the ground state absorption}

\begin{table}[H]
\centering
\begin{centering}
\caption{Calculated energies (in eV), transition dipole couplings with the
ground state, and wavefunction characteristics. $\boldsymbol{\mu}$
is the transition dipole coupling with the ground state (A\r{ }).
\textbf{CT$_{CC}$}, \textbf{CT$_{C\beta_{1}}$},\textbf{ CT$_{C\beta_{2}}$},
\textbf{CT$_{\beta\beta}$}, and \textbf{LE$_{\beta}$} are normalized
coefficients of configurations displaying chromophore to chromophore
CT, chromophore to nearest bridge CT (and vice versa), chromophore
to second nearest bridge CT(vice and versa), bridge to bridge CT and
localized excitation on the bridge, respectively.}
\par\end{centering}

\centering{}%
\begin{tabular}{|c|c|c|c|c|c|c|c|}
\hline 
Connectivity & E & $\mu$ & CT$_{CC}$ & CT$_{C\beta_{1}}$ & CT$_{C\beta_{2}}$ & CT$_{\beta\beta}$ & LE$_{\beta}$\tabularnewline
\hline 
\hline 
\multirow{4}{*}{\textit{anti-anti}} & 3.21 & 1.28 & 0 & 0.38 & 0 & 0 & 0.11\tabularnewline
\cline{2-8}
 & 3.36 & 0.97 & 0 & 0.14 & 0 & 0.54 & 0\tabularnewline
\cline{2-8}
 & 3.45 & 0.29 & 0 & 0.38 & 0 & 0.07 & 0\tabularnewline
\cline{2-8}
 & 3.52 & 0.47 & 0.53 & 0.08 & 0.19 & 0.16 & 0\tabularnewline
\hline 
\multirow{4}{*}{\textit{syn-syn}} & 3.21 & 1.43 & 0 & 0.38 & 0 & 0.14 & 0\tabularnewline
\cline{2-8}
 & 3.36 & 0.86 & 0 & 0.07 & 0.11 & 0.55 & 0.14\tabularnewline
\cline{2-8}
 & 3.45 & 0.33 & 0.10 & 0.13 & 0.11 & 0 & 0\tabularnewline
\cline{2-8}
 & 3.52 & 0.21 & 0.53 & 0 & 0.16 & 0.17 & 0\tabularnewline
\hline 
\multirow{7}{*}{\textit{syn-anti}} & \multirow{2}{*}{3.21} & 1.27 & 0 & 0.08 & 0.39, 0.37 & 0 & 0.13, 0.08\tabularnewline
\cline{3-8}
 &  & 0.47 & 0 & 0.06 & 0.40, 0.37 & 0.07 & 0.09, 0.08\tabularnewline
\cline{2-8}
 & 3.36 & 0.91 & 0.11 & 0.17, 0.15 & 0.06 & 0.53 & 0.14\tabularnewline
\cline{2-8}
 & \multirow{2}{*}{3.45} & 0.30 & 0.10 & 0.14 & 0.20 & 0.03 & 0\tabularnewline
\cline{3-8}
 &  & 0.25 & 0 & 0.06 & 0.23 & 0.05 & 0\tabularnewline
\cline{2-8}
 & 3.52 & 0.27 & 0.53 & 0.14, 0.18 & 0.11, 0.05 & 0.05 & 0.05\tabularnewline
\cline{2-8}
 & 3.63 & 0.22 & 0 & 0.43 & 0.09, 0.08 & 0 & 0\tabularnewline
\hline 
\end{tabular}
\end{table}

\section{Charge transfer states in the Excited state absorption}

\begin{table}[H]
\centering
\begin{centering}
\caption{Calculated energies (in eV), transition dipole couplings with the
optical singlet state, and wavefunction characteristics.\textbf{ $\mu$}
is the transition dipole coupling with the optical singlet state S1.
Where \textbf{CT$_{CC}$}, \textbf{CT$_{C\beta_{1}}$}, \textbf{CT$_{C\beta_{2}}$},
\textbf{CT$_{\beta\beta}$}, and \textbf{LE$_{\beta}$} are normalized
coefficients of configurations displaying chromophore to chromophore
CT, chromophore to nearest bridge CT (and vice versa), chromophore
to second nearest bridge CT (and vice versa), bridge to bridge CT
and localized excitation on the bridge, respectively. The complete
WFs for the state in \textit{anti-anti}-\textbf{P-T-T-P} at 3.33 eV,
in \textit{syn-syn} \textbf{P-T-T-P} at 3.32 eV, and in \textit{syn-anti}
at 3.30 eV are dominated by intramonomer HOMO-1 to LUMO/HOMO to LUMO+1
excitation, not charge transfer.}
\par\end{centering}
\centering{}%
\begin{tabular}{|c|c|c|c|c|c|c|c|}
\hline 
Connectivity & E & $\mu$ & CT$_{CC}$ & CT$_{C\beta_{1}}$ & CT$_{C\beta_{2}}$ & CT$_{\beta\beta}$ & LE$_{\beta}$\tabularnewline
\hline 
\hline 
\multirow{4}{*}{\textit{anti-anti}} & 3.17 & 2.62 & 0 & 0.35 & 0.06 & 0 & 0\tabularnewline
\cline{2-8}
 & 3.33 & 0.36 & 0 & 0 & 0.13 & 0.53 & 0\tabularnewline
\cline{2-8}
 & 3.34 & 2.21 & 0 & 0.19 & 0 & 0.11 & 0\tabularnewline
\cline{2-8}
 & 3.51 & 0.13 & 0.53 & 0.06 & 0.20 & 0.14 & 0\tabularnewline
\hline 
\multirow{3}{*}{\textit{syn-syn}} & 3.17 & 2.47 & 0 & 0.35 & 0 & 0.10 & 0\tabularnewline
\cline{2-8}
 & 3.32 & 2.07 & 0 & 0.09 & 0.07 & 0.35 & 0\tabularnewline
\cline{2-8}
 & 3.37 & 1.21 & 0.04 & 0.07 & 0.11 & 0.54 & 0.14\tabularnewline
\hline 
\multirow{4}{*}{\textit{syn-anti}} & $3.14$ & 2.34 & 0 & 0.51 & 0.09 & 0.07 & 0\tabularnewline
\cline{2-8}
 & \multirow{1}{*}{$3.16$} & 0.17 & 0 & 0.51 & 0.09 & 0 & 0\tabularnewline
\cline{2-8}
 & 3.30 & 2.25 & 0.06 & 0.14 & 0.12, 0.07 & 0.28 & 0\tabularnewline
\cline{2-8}
 & 3.34 & 1.09 & 0.05 & 0.19 & 0.11 & 0.46 & 0\tabularnewline
\hline 
\end{tabular}
\end{table}

\section{Normalized overall spin-singlet triplet-triplet wavefunctions of
syn-syn PT2P}

\begin{figure}[H]
\centering{}\includegraphics[scale=0.22]{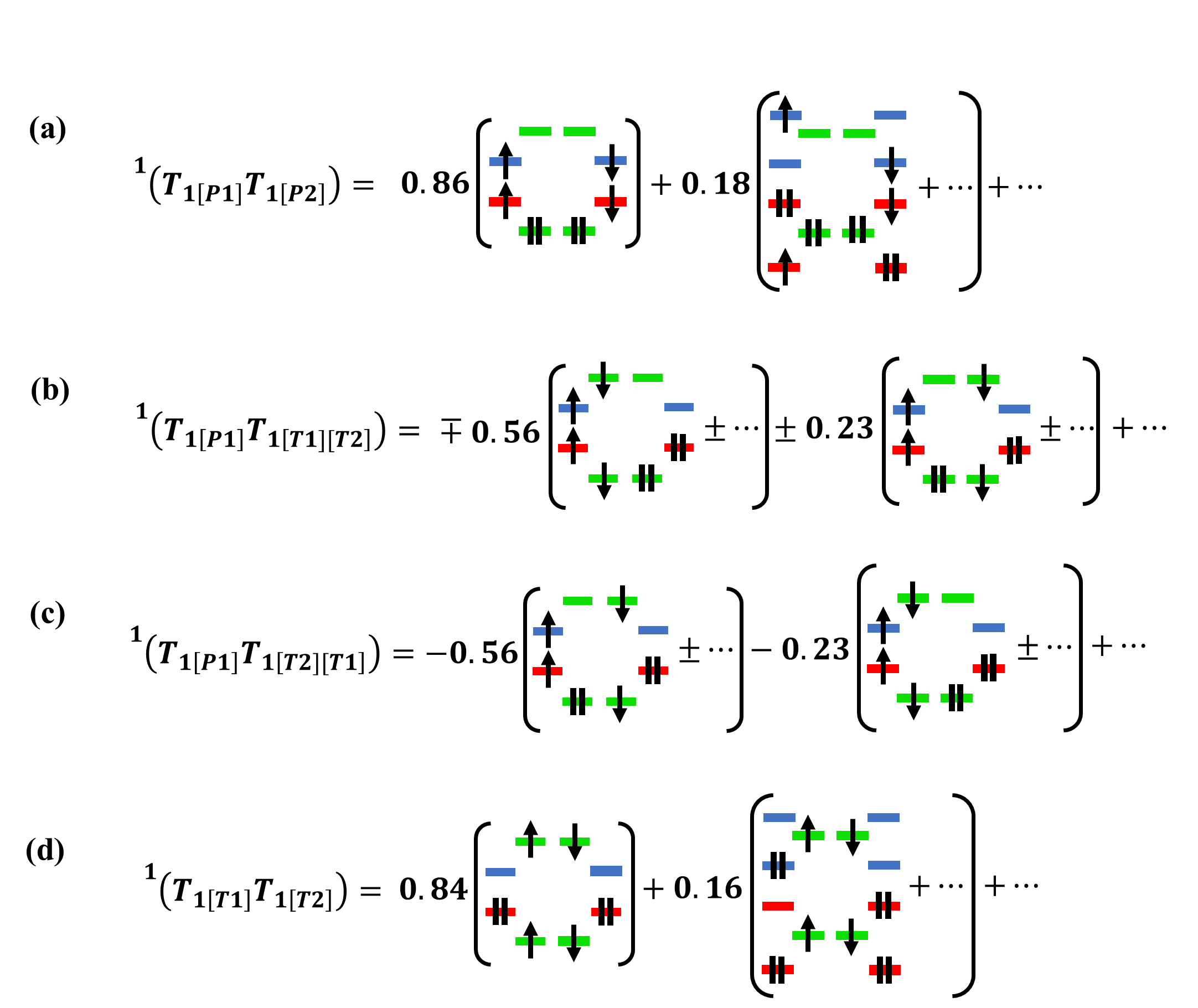}\caption{Normalized wavefunctions of (a) $^{1}$(T$_{1[P_{1}]}$T$_{1[P_{2}]}$),
(b)$^{1}$(T$_{1[P_{1}]}$T$_{1[T\text{1}||T2]}$), (c)$^{1}$(T$_{1[P_{1}]}$T$_{1[T\text{2}||T1]}$)
and (d)$^{1}$(T$_{1[T1]}$T$_{1[T2]}$) states in \textit{syn-syn}-\textbf{P-T-T-P}.}
\end{figure}

\section{Normalized overall spin-singlet triplet-triplet wavefunctions of
syn-anti PT2P}

\begin{figure}[H]
\centering{}\includegraphics[scale=0.22]{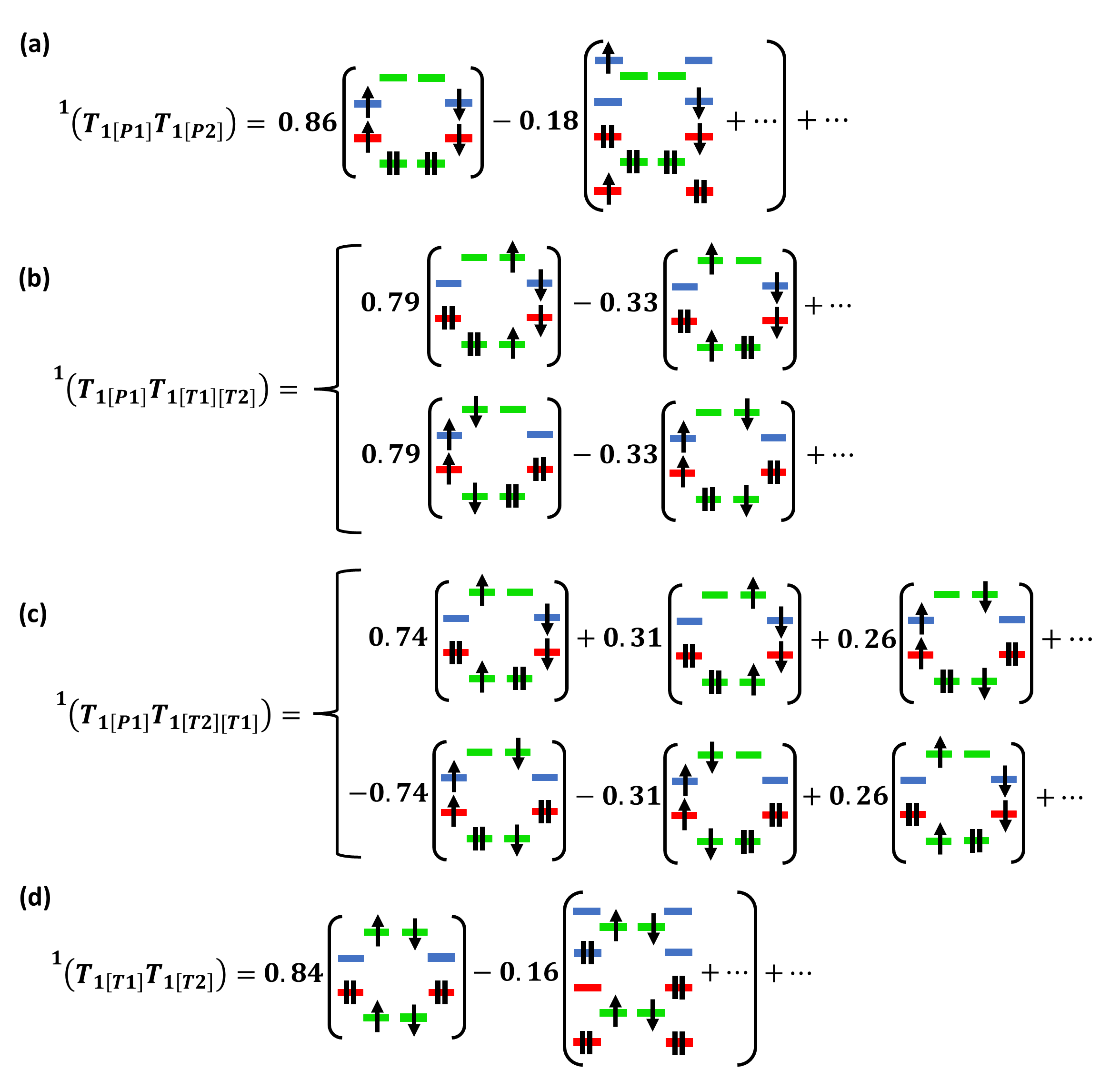}\caption{Normalized wavefunctions of (a) $^{1}$(T$_{1[P1]}$T$_{1[P2]}$),
(b)$^{1}$(T$_{1[P1]}$T$_{1[T1][T2]}$), (c)$^{1}$(T$_{1[P1]}$T$_{1[T2][T1]}$)
and (d)$^{1}$(T$_{1[T1]}$T$_{1[T2]}$) states in \textit{syn-anti}-\textbf{P-T-T-P}.
Where both (b) and (c) are doubly degenerate triplet-triplet states.}
\end{figure}

\newpage{}

\section{Energy of triplet-triplet states in syn-syn and syn-anti PT2P}

\begin{table}[H]
\caption{Calculated energies (in eV) of various triplet-triplet states for
syn-syn and syn-anti\textbf{ P-T2-P}.}

\centering{}%
\begin{tabular}{ccc}
\toprule 
\multirow{2}{*}{State} & \multicolumn{2}{c}{Energy}\tabularnewline
\cmidrule{2-3}
 & syn-syn & syn-anti\tabularnewline
\midrule 
$^{1}$(T$_{1[P1]}$T$_{1[P2]}$) & 2.08 & 2.08\tabularnewline
\midrule 
$^{1}$(T$_{1[P1]}$T$_{1[T1][T2]}$) & 2.42 & 2.45\tabularnewline
\midrule 
$^{1}$(T$_{1[P1]}$T$_{1[T2][T1]}$) & 2.45 & 2.47\tabularnewline
\midrule 
$^{1}$(T$_{1[T1]}$T$_{1[T2]}$) & 2.77 & 2.77\tabularnewline
\bottomrule
\end{tabular}
\end{table}